# Fundamental Limits of Exciton-Exciton Annihilation for Light Emission in Transition Metal Dichalcogenide Monolayers

Yiling Yu[1,2§], Yifei Yu[1§], Chao Xu[2§], Andy Barrette[2], Kenan Gundogdu[2]*, Linyou Cao[1,2]*

[1]Department of Materials Science and Engineering, North Carolina State University, Raleigh NC 27695; [2]Department of Physics, North Carolina State University, Raleigh NC 27695;

[§] These authors contribute equally.

**Abstract**

We quantitatively illustrate the fundamental limit that exciton-exciton annihilation (EEA) may impose to the light emission of monolayer transition metal dichalcogenide (TMDC) materials. The EEA in TMDC monolayers shows dependence on the interaction with substrates as its rate increases from 0.1 cm$^2$/s (0.05 cm$^2$/s) to 0.3 cm$^2$/s (0.1 cm$^2$/s) with the substrates removed for WS$_2$ (MoS$_2$) monolayers. It turns to be the major pathway of exciton decay and dominates the luminescence efficiency when the exciton density is beyond $10^{10}$ cm$^{-2}$ in suspended monolayers or $10^{11}$ cm$^{-2}$ in supported monolayers. This sets an upper limit on the density of injected charges in light emission devices for the realization of optimal luminescence efficiency. The strong EEA rate also dictates the pumping threshold for population inversion in the monolayers to be 12-18 MW/cm$^2$ (optically) or 2.5-4×$10^5$ A/cm$^2$ (electrically).

* To whom correspondence should be addressed

Email: lcao2@ncsu.edu, kgundog@ncsu.edu

Two-dimensional (2D) transition metal dichalcogenide (TMDC) materials such as monolayer $MoS_2$ and $WS_2$ promise to enable the development of atomic-scale light emission devices owing to their semiconducting nature, perfect surface passivation, and strong exciton binding energy [1]. A key issue for the device development is to understand the exciton dynamics of these materials, which has been known bearing substantial difference from what observed at conventional materials. In particular, the extraordinary exciton binding energy in the TMDC monolayers [2-5] is expected to enable strong many-body interactions like exciton-exciton annihilation (EEA). Recent studies have demonstrated that the EEA rate in monolayer TMDC materials is indeed two orders of magnitude higher than that in conventional semiconductor materials [6-9]. However, much fundamental of the EEA has remained to be elusive. For instance, substantial discrepancy can be found in the previous studies as some reported negligible EEA in the monolayers[10-12] shown to have strong EEA by others[7,8]. It is also not clear how the EEA could depend on the nature of the materials and the environment at the proximity like substrates. Most importantly, although it is generally known that EEA may affect luminescence efficiency, there is no quantitatively understanding about how the strong EEA could affect the light emission efficiency of the monolayers in unusual ways. This understanding would provide useful guidance for the rational design of high-performance light-emission devices.

Here we quantitatively elucidate the fundamental limit that the strong EEA may impose to the luminescence efficiency and lasing threshold in monolayer TMDC materials. We evaluate the EEA and its effect on luminescence for both suspended monolayers and monolayers supported by substrates. The EEA is found subject to influence of substrates as the substrate may decrease the EEA rate and facilitates defect-assisted recombination that can compete with the EEA as the pathway for excitons to decay. The EEA may turn to be the major decay pathway and dominate the luminescence efficiency when the density of excitons is in scale of $10^{10}$ cm$^{-2}$ at suspended monolayers or $10^{11}$ cm$^{-2}$ at supported monolayers. This sets an upper limit on the density of injected charges in light emission devices in order to achieve optimal luminescence efficiency. The strong EEA also dictates the pumping threshold for population inversion in the monolayers to be 12-18 MW/cm$^2$ at optically pumping or 2.5-4×10$^5$ A/cm$^2$ at electrically pumping.

Fig. 1a-b shows the PL efficiencies (the number of emitted photons vs. the number of adsorbed photons) of suspended monolayer MoS$_2$ and WS$_2$ as a function of incident laser power. The samples were prepared by manually transferring chemical vapor deposition-grown monolayers from the growth substrate (sapphire) onto SiO$_2$/Si substrates pre-patterned with holes (see Methods and Fig. S1 in the Supplemental Materials)[13]. The efficiency is evaluated from PL measurements at room temperature with Rhodamine 6G used as a reference. While the efficiencies vary among these materials, all exhibit an exponential decrease with the incident power increasing, even at an incident power as low as 10 W/cm$^2$. In stark contrast, the PL efficiencies of the as-grown monolayers show much milder dependence on the incident power (Fig. 1c-d). We can exclude out any substantial heating effects and the formation of bi-excitons in the measurement as the lineshape and position of the PL show negligible change through the measurement (Fig. S2). It has been known that the PL would redshift or broaden at elevated temperatures and show new peaks at lower energy with the formation of bi-excitons [14-19]. With the exclusion of heating effects and biexcitons, the observed power-dependent PL efficiency may be correlated to another non-linear process: exciton-exciton annihilation (EEA).

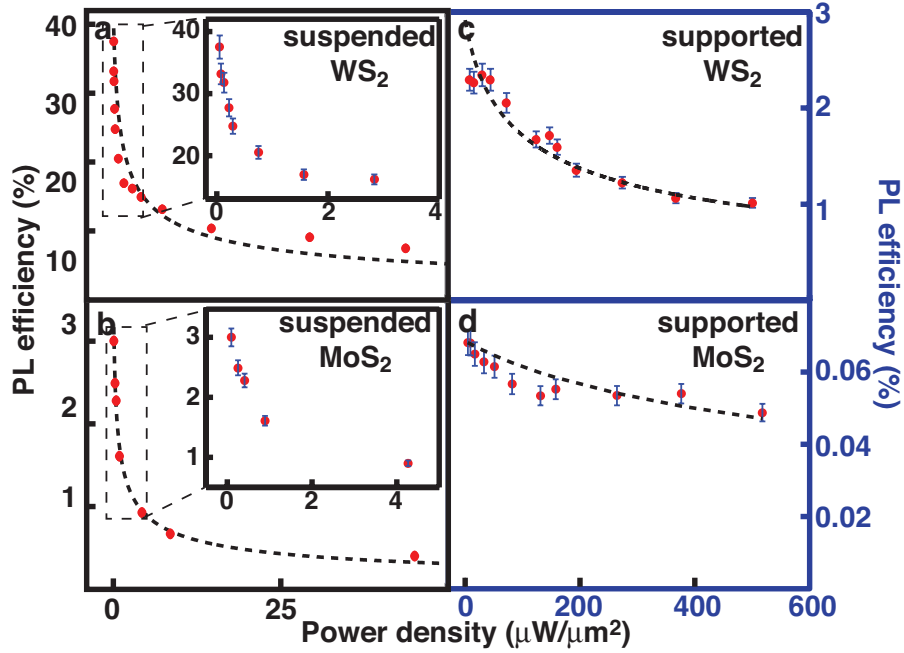

**FIG 1.** PL efficiencies of (a) suspended monolayer $WS_2$, (b) suspended monolayer $MoS_2$, (c) as-grown monolayer $WS_2$ on sapphire substrates, and (d) as-grown monolayer $MoS_2$ on sapphire substrates as a function of the incident power density. The dashed lines are simulation results using eq. (3) and the parameters given in Table 1. The insets in (a) and (b) are to better illustrate the results in the corresponding dashed box. All the given error bars are 10%. The error bars in the (a) and (b) are ignored for visual convenience.

To better understand the EEA, we examined the exciton dynamics in the suspended monolayers using pump-probe techniques (see Methods). What we measured is the differential reflection $\Delta R/R$ of a delayed probe beam from the monolayers photoexcited by a pump beam (590 nm). The wavelength of the probe beam is chosen to match the *A* exciton of the monolayer, and the pumping fluence is set to be small enough to ensure the absorption far below saturation. As a result, the differential reflection ($\Delta R/R$) can be linearly correlated to the density of photo-generated charge carriers at the band edges. Fig. 2 shows the transient differential reflection $\Delta R/R$ collected from suspended $WS_2$ monolayers (see Fig. S3 for the result of suspended $MoS_2$ monolayers). We confirmed no substantial heating effect in the experiments by ensuring a reasonable linear dependence of the $\Delta R/R$ at the 0s delay $(\Delta R/R)_0$ on the pumping fluence because $(\Delta R/R)_0$ is sensitive to the temperature (Fig. S6). The decay rate can be found increasing with the pumping fluence (Fig. 2a), consistent with what expected from EEA. The increase of the decay rate also indicates negligible formation of bi-excitons, which would otherwise show the decay rate slowing with the pump fluence increasing [19].

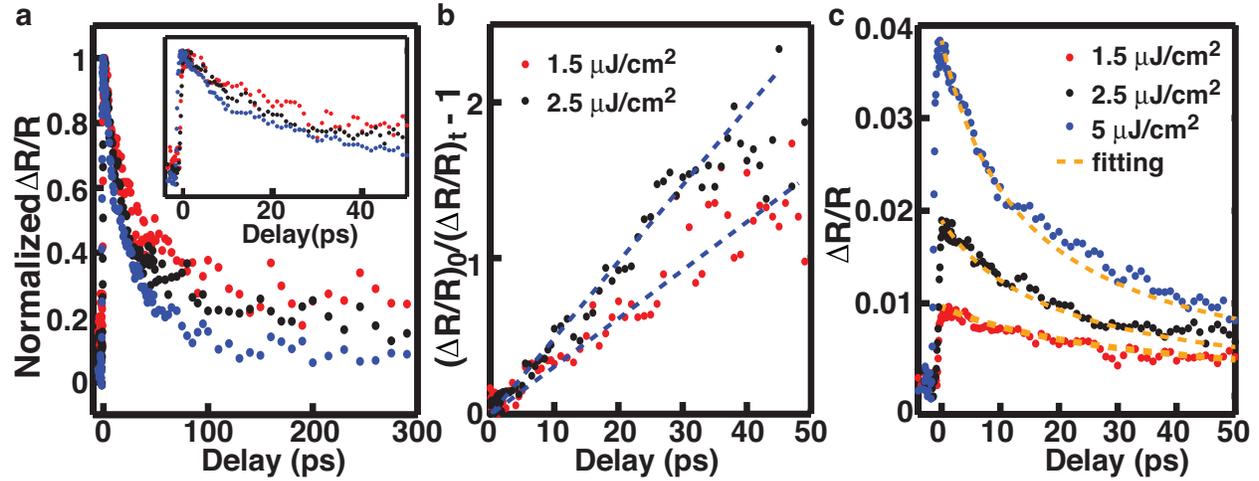

**FIG 2.** (a) Normalized differential reflection of suspended WS$_2$ with different pumping fluences, 1.5 µJ/cm$^2$ (red), 2.5 µJ/cm$^2$ (blue), and 5.0 µJ/cm$^2$ (black). Inset: the results for the early stage of the decay. (b) The result of $(\Delta R/R)_0/(\Delta R/R)_t$ -1 derived from the data in (a). The dashed line serves to illustrate the slope of the result. (c) Fitting for the measured differential reflection of suspended WS$_2$ with different pumping fluences as labeled, The fitted results are plotted in dashed lines and the experimental results are dots.

We can evaluate the rate constant of the EEA based on the pump-probe measurement. Should the exciton decay be dominated by EEA, the rate equation of exciton density would be written as a function of the EEA rate $k_{ee}$, $dN/dt = -k_{ee}N^2$. And the exciton density $N(t)$ would be correlated to the total photo-generated excitons $N_0$ as

$$\frac{N_0}{N(t)} - 1 = k_{ee} N_0 t \quad (1)$$

As $\Delta R/R$ can be linearly correlated to the density of photo-generated charge carriers, we may have $N_0/N(t) = (\Delta R/R)_0/(\Delta R/R)_t$. We can derive $(\Delta R/R)_0/(\Delta R/R)_t$ -1 from the result given in Fig. 2a, and plot it as a function of the delay time in Fig. 2b. The result shows that $(\Delta R/R)_0/(\Delta R/R)_t$ -1 linearly depends on the delay time at the early stage of the decay (up to 50-100 ps) and its slope linearly increases with the pumping fluence (Fig. 2b). This is consistent with what expected from eq. (1), indicating that the early-stage exciton decay in the suspended monolayer is dominated by EEA. We can also estimate the total photo-generated excitons $N_0$ from the incident fluence and the absorption efficiency of the monolayers. The absorption of suspended $WS_2$ and $MoS_2$ for the pump beam is estimated to be 0.058 and 0.022, respectively, using the refractive index we measured (See Ref. 3 and Fig. S7). The rate constant $k_{ee}$ can thus be derived from the slope in Fig. 2b as 0.3 cm$^2$/s and 0.1 cm$^2$/s for suspended $WS_2$ and $MoS_2$ monolayers, respectively.

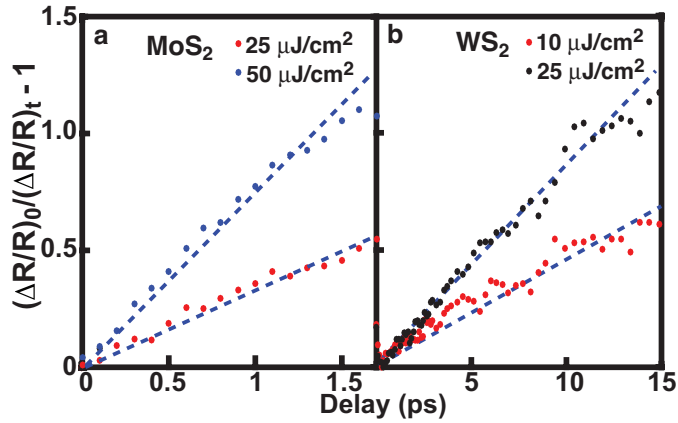

**FIG 2.** $(\Delta R/R)_0/(\Delta R/R)_t$ -1 of (a) as-grown $MoS_2$ and (b) as-grown $WS_2$. The results are derived from the differential reflection measurement at these materials with different pumping fluences as labeled. The differential reflection measurement results are given in Fig. S4-S5.

To understand the different power dependence of PL efficiency in the supported monolayers, we performed similar pump-probe measurements and data analysis for the as-grown $MoS_2$ and $WS_2$ monolayers onto sapphire substrates (Fig. 3 and Fig.S4-S5). The EEA rate is found to be 0.1 cm$^2$/s and 0.05 cm$^2$/s for the supported $WS_2$ and $MoS_2$, respectively. This smaller EEA rate indicates the effect of substrates, which may be understood from an intuitive perspective. Generally, the rate of EEA is related with the diffusion coefficient of excitons $D$ and the annihilation radius $R$ that represents the separation of two excitons when the annihilation may occur, $k_{ee} = 4\pi DR$ [20]. The presence of substrates may lower charge mobility and hence the diffusion coefficient [21]. The substrate may also lower the exciton binding energy[22,23], which could subsequently lead to a smaller *R*. Additionally, the substrate may facilitate defect-assisted recombination that can compete with the EEA as a pathway for excitons to decay [11,24]. While the presence of defect-assisted recombination may not change the EEA rate, it could make the experimental observation of the EEA more difficult, particularly when the defect-assisted decay rate is comparable to or even faster than the EEA rate. For instance, the EEA in the as-grown MoS$_2$ can be observed only in the first several ps (< 2 ps) and with relatively high pumping fluence (> 25 µJ/cm$^2$) (Fig.3a and Fig. S5). We found in experiments that generally it was generally more difficult to observe the EEA in the monolayers showing lower PL intensities. Given the significant effect of substrates on the EEA, we believe that the discrepancies in the previous studies, i.e., the demonstration of different EEA rates in the same materials [7,8,10-12], is likely due to difference in the effect of substrates.

We can better understand the effect of the EEA on luminescence efficiency (Fig.1) by correlating the power-dependent efficiency to the nonlinear and linear decay processes involved. The rate equation of exciton density for the time-averaged PL can be written as

$$\frac{dN}{dt} = -(\frac{1}{\tau_r} + \frac{1}{\tau_{nr}})N - k_{ee}N^2 + \alpha I_0 \quad (2)$$

where $\tau_r$ and $\tau_{nr}$ represent the exciton lifetimes associated with radiative and linear non-radiative recombinations, $\alpha$ and $I_0$ are the absorption efficiency for the incident wavelength and the incident power density. From eq. (2) we can derive the efficiency of the time-averaged PL as

$$QY = \frac{N/\tau_r}{\alpha I_0} = \frac{\left[\sqrt{(1/\tau_r + 1/\tau_{nr})^2 + 4k_{ee}\alpha I_0} - (1/\tau_r + 1/\tau_{nr})\right]}{2k_{ee}\alpha I_0 \tau_r} \quad (3)$$

The absorption efficiency $\alpha$ of suspended monolayer $MoS_2$ and $WS_2$ can be calculated using the refractive index we measured (See Ref. 3 and Fig. S7), which is 0.065 and 0.055 for the incidence of 532 nm, respectively. The EEA rate $k_{ee}$ is known from the differential reflection measurement. Then we can evaluate $\tau_r$ and $\tau_{nr}$ by numerically fitting the measured power-dependent efficiency to eq. (3). The fitting results are plotted (dashed lines) along with the experimental results in Fig. 1 and the fitted value of $\tau_r$ and $\tau_{nr}$ are given in Table 1.

Table 1. EEA rate and lifetime

|  | $k_{ee}$ (cm²/s) | $\tau_r$ (ns) | $\tau_{nr}$ (ns) |
|---|---|---|---|
| **Sus WS₂** | 0.3 | 1 | 0.76 |
| **AG WS₂** | 0.1 | 4.5 | 0.13 |
| **Sus MoS₂** | 0.1 | 28 | 1 |
| **AG MoS₂** | 0.05 | 80 | 0.05 |

This result may provide useful guidance for the rational design of light emission devices with optimal efficiency. It can guide the proper charge injection in light emission devices for the realization of optimal quantum yield. According to eq. (2), the EEA may turn to be the major pathway of exciton decay ($k_{ee}N^2 > (1/\tau_r + 1/\tau_{nr})N$) when the exciton density $N > (1/\tau_r + 1/\tau_{nr})/k_{ee}$, which is in the scale of $10^{10}$ cm$^{-2}$ and $10^{11}$ cm$^{-2}$ for suspended and supported monolayers, respectively. The charge injection must be controlled to maintain the steady-state charge density well below those values. We can estimate the steady-state charge density as a function of the injected current density $J$ using an equation modified from eq.(2) $N = \left[\sqrt{(1/\tau_r + 1/\tau_{nr})^2 + 4k_{ee}J} - (1/\tau_r + 1/\tau_{nr})\right]/2k_{ee}$. The calculation indicates that, in order to maintain the steady-state charge density well below (<10%) of the threshold values, the injected current density should be no more than 0.2-0.4 A/cm² and 15-30 A/cm² for suspended and supported monolayers, respectively (Fig. 4a). Additionally, the result may help predict the lasing threshold and optical gain coefficient. We use a simple three-level system to represent the pumping process in the monolayer (Fig.4b inset), in which the charges at the ground state 1 (valence band edge) are first pumped to the upper pump level 3 (a higher level in the conduction band) and then quickly decay to the level 2 (conduction band edge). Our analysis indicates the population inversion is completely dictated by the EEA with negligible influence from the linear recombination processes $\tau_r$ and $\tau_{nr}$ (see S1 in the Supplemental Materials) as

$$\Delta N = \left(\sqrt{W_p^2 + 4k_{ee}W_pN_t} - W_p\right)/k_{ee} - N_t \qquad (4)$$

and the optical gain coefficient as $\gamma = \sigma_{12}\Delta N$, where $W_p = \sigma_{13}I_p/h\nu_{13}$ representing the pumping rate. The total charge density $N_t$ in monolayer $WS_2$ and $MoS_2$ can be estimated to be $4.17\times10^{12}$ cm$^{-2}$ and $6.27\times10^{12}$ cm$^{-2}$ by assuming parabolic band edges at $K$ point and using the average effective mass reported in the literature ($0.4m_0$ and $0.6m_0$ for $WS_2$ and $MoS_2$)[25-28]. The stimulated emission (absorption) cross-section $\sigma_{13}$ ($\sigma_{12}$) can be derived from the total charge density and the absorption efficiency $\alpha_{13}$ ($\alpha_{12}$) as $\sigma_{13} = \alpha_{13}/N_t$ ($\sigma_{12} = \alpha_{12}/N_t$). Without losing generality, we use the pumping wavelength of 532 nm as an example toimplement numerical evaluation. Fig. 4b-c shows the calculated population inversion and optical gain coefficient as a function of the incident power at optically pumping (532 nm) and the injected current density at electrically pumping. The result indicates that the threshold is around 12-18 MW/cm$^2$ at optically pumping or 2.5-4 MA/cm$^2$ at electrically pumping. This calculation does not take into account any optical enhancement effects, heating effect during the pumping, and possible re-normalization of the bandgap [23,29]. It nevertheless provides useful guidance for the development of 2D TMDC lasers operated at room temperatures. This predicted threshold pumping power is reasonably consistent with one recent study, in which the threshold pumping power for lasing in supported $WS_2$ monolayer is estimated at 5-8 MW/cm$^2$ [31].

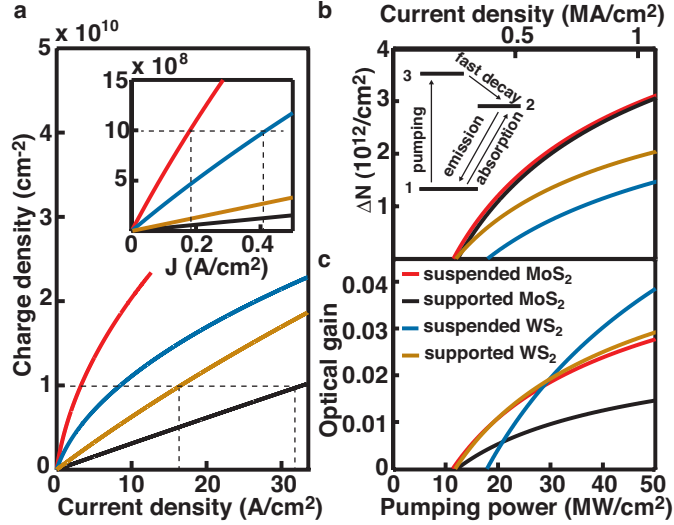

FIG. 4. (a) Steady-state charge density as a function of the injected current density in suspended monolayer $MoS_2$ (red), supported $MoS_2$ (black), suspended $WS_2$ (blue), and supported $WS_2$ (brown). The dash lines indicate the proper charge density and corresponding injection current density in order to have negligible effects from the EEA. (b) Calculated population inversion and (c) optical gain coefficients as a function of pumping power (optical) and injection current density (electrical) for different monolayers, including suspended monolayer $MoS_2$ (red), supported $MoS_2$ (black), suspended $WS_2$ (blue), and supported $WS_2$ (brown). Inset, a schematic illustration of the three-level model used for the calculation. The absorption efficiency is approximately set to be 5% for the conversion of the pumping power to the injection current density.

In conclusion, we have quantitatively evaluated the EEA and its effect on light emission for suspended and supported monolayer TMDC materials. The EEA is subject to strong influence of the substrate. It may turn to be the major pathway of exciton decay and dominates the luminescence efficiency when the exciton density is in scale of $10^{10}$ cm$^{-2}$ in suspended monolayers or $10^{11}$ cm$^{-2}$ in suspended monolayers. This sets an upper limit for the density of injected charges in light emission devices in order to achieve optimal luminescence efficiency. The strong EEA also dictates the pumping threshold for population inversion in the monolayers to be 12-18 MW/cm$^2$ at optically pumping or 2.5-4 MA/cm$^2$ at electrically pumping. The result may provide useful guidance for the rational design of atomic-scale light emission devices, including LEDs and lasers.

[30] Y. Ye, Z. J. Wong, X. Lu, X. Ni, H. Zhu, X. Chen, Y. Wang, and X. Zhang, Nat. Photon. **9**, 733 (2015).

*Supplemental Materials*

# Fundamental Limits of Exciton-Exciton Annihilation for Light Emission in Transition Metal Dichalcogenide Monolayers

Yiling Yu[1,2,§], Yifei Yu[1,§], Chao Xu[2,§], Andy Barrette[2], Kenan Gundogdu[2]*, Linyou Cao[1,2]*

[1]Department of Materials Science and Engineering, North Carolina State University, Raleigh NC 27695;
[2]Department of Physics, North Carolina State University, Raleigh NC 27695;

[§] These authors contribute equally.

* To whom correspondence should be addressed.
Email: lcao2@ncsu.edu, kgundog@ncsu.edu

This PDF document includes

Figure S1-S7

S1. Derivative for the population inversion in the monolayers

References

**Methods**

*Synthesis and transfer of MoS$_2$, WS$_2$, and WSe$_2$ monolayers:* The monolayers were grown using a chemical vapor deposition (CVD) reported previously[1]. Typically, 1g sulfur or selenium powder (Sigma-Aldrich) and 15-30mg MoO$_3$ (WO$_3$) (99.99%, Sigma-Aldrich) source material were placed in the upstream and the center of a tube furnace, respectively. And substrates (usually sapphire) were placed at the downstream of the tube. Typical growth was performed at 750-900 °C) for 10 (30) minutes under a flow of Ar gas in rate of 100 sccm and ambient pressure.

The transfer of the monolayers followed a surface-energy-assisted transfer approach that we have developed previously[2]. In a typical transfer process, 9 g of polystyrene (PS) with a molecular weight of 280000 g/mol was dissolved in 100 mL of toluene, and then the PS solution was spin-coated (3000 rpm for 60 s) on the as-grown monolayers. This was followed by a baking at 80−90 °C for 1 hour. A water droplet was then dropped on top of the monolayer. Due to the different surface energies of the monolayer and the substrate, water molecules could penetrate under the monolayer, resulting the delamination of the PS-monolayer assembly. We could pick up the polymer/monolayer assembly with a tweezers and transferred it to different substrates. After that, we baked the transferred PS-monolayer assembly at 80 °C for 1 h and performed a final baking for 30 min at 150 °C. Finally, PS was removed by rinsing with toluene several times.

*Characterizations:* Raman and AFM measurements were used to confirm that the synthesized samples are monolayers. The Raman measurements were carried out by Horiba Labram HR800 system with a 532 nm laser. AFM measurements were performed at a Veeco Dimension-3000 atomic force microscope. A home-built setup that consists of a confocal microscope (Nikon Eclipse C1) connected with a monochromator (SpectraPro, Princeton Instruments) and a detector (Pixis, Princeton Instruments) was used to perform the photoluminescence measurement with an excitation wavelength of 532nm.

A 150 fs pulse at 2.10 eV is used to pump electrons from the valence band into conduction band of the monolayers. The differential reflection ($\Delta R/R$) of a time-delayed probe pulse, whose wavelength is chosen to match the A exciton transition (~1.88eV for MoS$_2$ and 2eV for WS$_2$, respectively), was used to probe the excitation dynamics. The pump and probe beams were collinearly polarized and co-focused using a 50X long working distance objective and the reflected probe pulse was collected using the same objective. The size of the focused beam is about 2μm. A monochromotor and a Si photodetector combination measure the differential reflection using lock-in amplification method. Unless otherwise specified, all experiments were performed at room temperature.

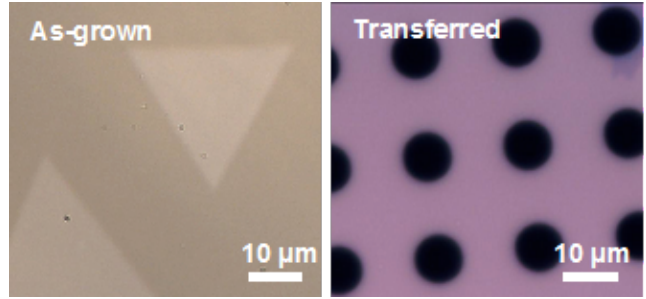

**Figure S1.** Optical image of typical as-grown monolayers on sapphire substrates and transferred monolayers on SiO$_2$/Si substrates pre-patterned with holes. The black areas are the pre-patterned holes on the substrate.

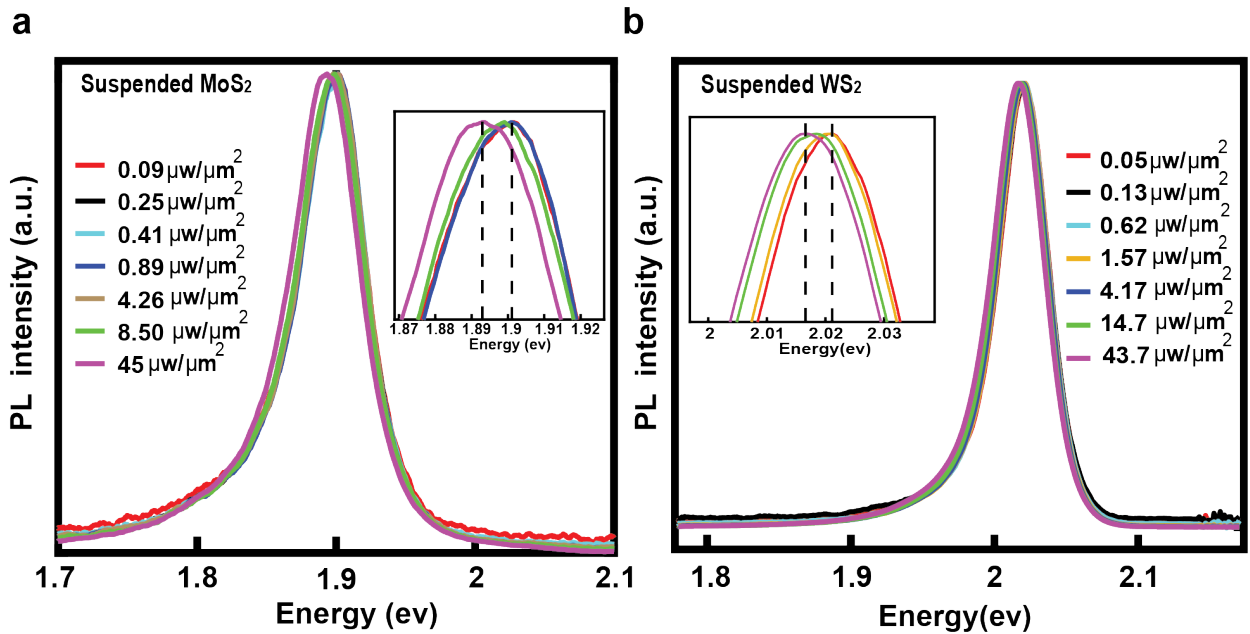

**Figure S2. Dependence of the PL spectra of suspended MoS$_2$ and WS$_2$ on incident power.** (a) PL spectra of suspended MoS$_2$ at different incident powers. Inset: Zoom-in curves of 4 representative powers, the black dash lines denotes PL peak position at the lowest and highest incident powers used in the experiments. The temperature change in the sample can be estimated by this peak shift (~ 8 meV) using the well-established temperature dependent bang-gap equation $E_g(T) = E_g(0) - \alpha T^2 / (T + \beta)$ and $\alpha = 5.9 \times 10^{-4} eV/K$, $\beta = 430K$ from Ref.S3, the temperature change is calculated to be around 20K with the indicent power 50 μW/cm$^2$. (b) PL spectra of suspended WS$_2$ at different incident powers. Similar to suspended MoS$_2$, the peak shift is small and there is essentially no change in the PL lineshape.

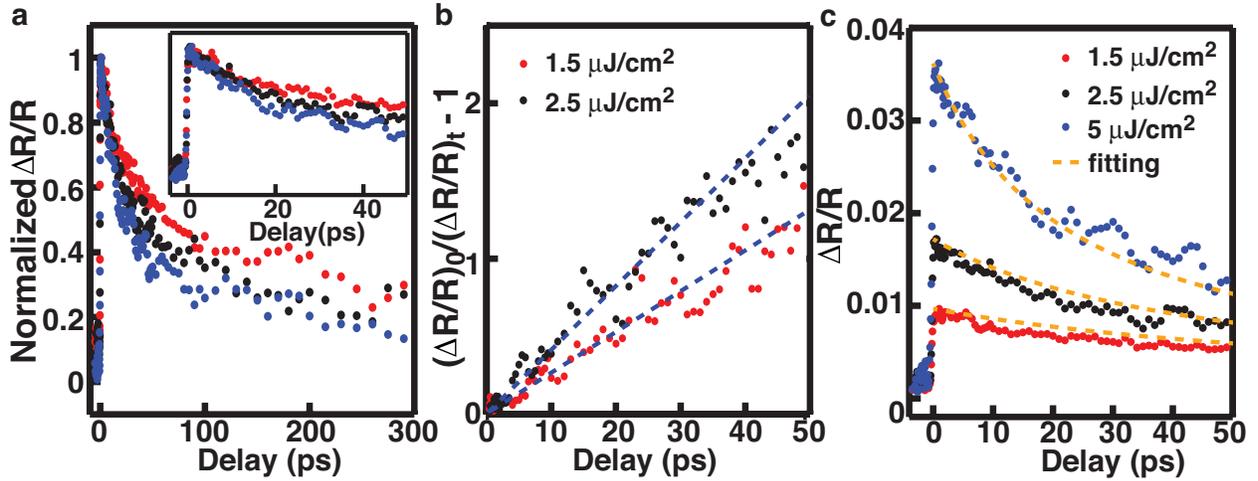

**Figure S3.** (a) Normalized differential reflection of suspended MoS$_2$ as a function of the time delay and with different pumping fluences, 1.5 µJ/cm$^2$ (red), 2.5 µJ/cm$^2$ (black), and 5.0 µJ/cm$^2$ (blue). Inset: the results for the early stage of the decay. (b) The result of $(\Delta R/R)_0/(\Delta R/R)_t -1$ derived from the data in (a) as a function of the delay time. The result for the pumping fluence of 5.0 µJ/cm$^2$ is not shown for the visual convenience. (c) Fitting for the measured differential reflection of suspended MoS$_2$ with different pumping fluences, The fitted results are plotted in dashed lines and the experimental results are dots, 1.5 µJ/cm$^2$ (red), 2.5 µJ/cm$^2$ (black), and 5.0 µJ/cm$^2$ (blue).

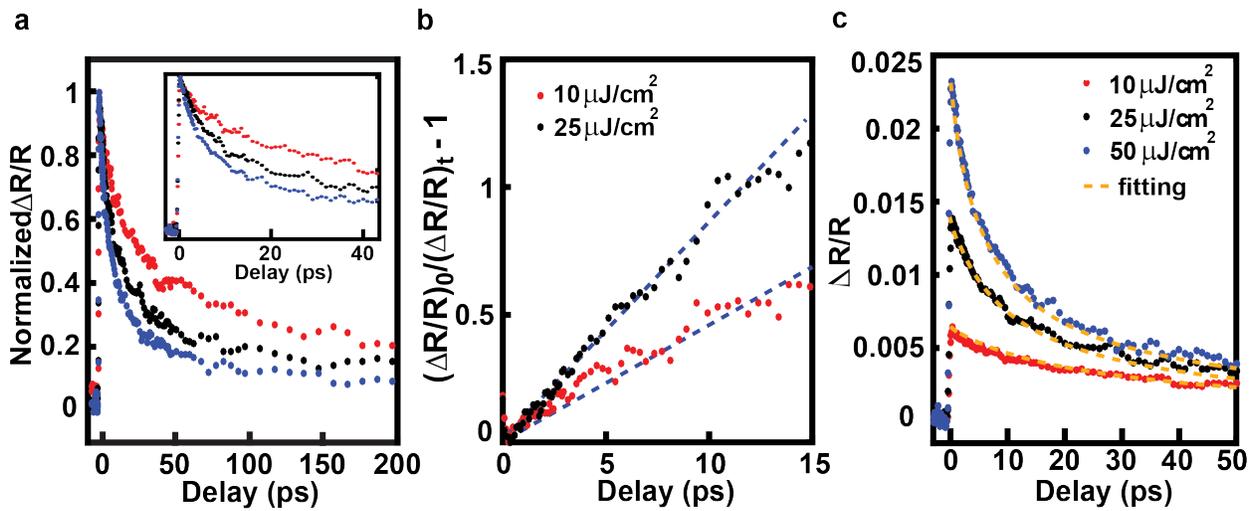

**Figure S4.** (a) Normalized differential reflection of as-grown WS$_2$ as a function of the time delay and with different pumping fluences, 10 µJ/cm$^2$ (red), 25 µJ/cm$^2$ (black), and 50 µJ/cm$^2$ (blue). Inset: the results for the early stage of the decay. (b) The result of $(\Delta R/R)_0/(\Delta R/R)_t -1$ derived from the data in (a) as a function of the delay time. The result for the pumping fluence of 50 µJ/cm$^2$ is not shown for the visual convenience. (c) Fitting for the measured differential reflection of as grown WS$_2$ with different pumping fluences, The fitted results are plotted in dashed lines and the experimental results are dots, 10 µJ/cm$^2$ (red), 25 µJ/cm$^2$ (black), and 50 µJ/cm$^2$ (blue).

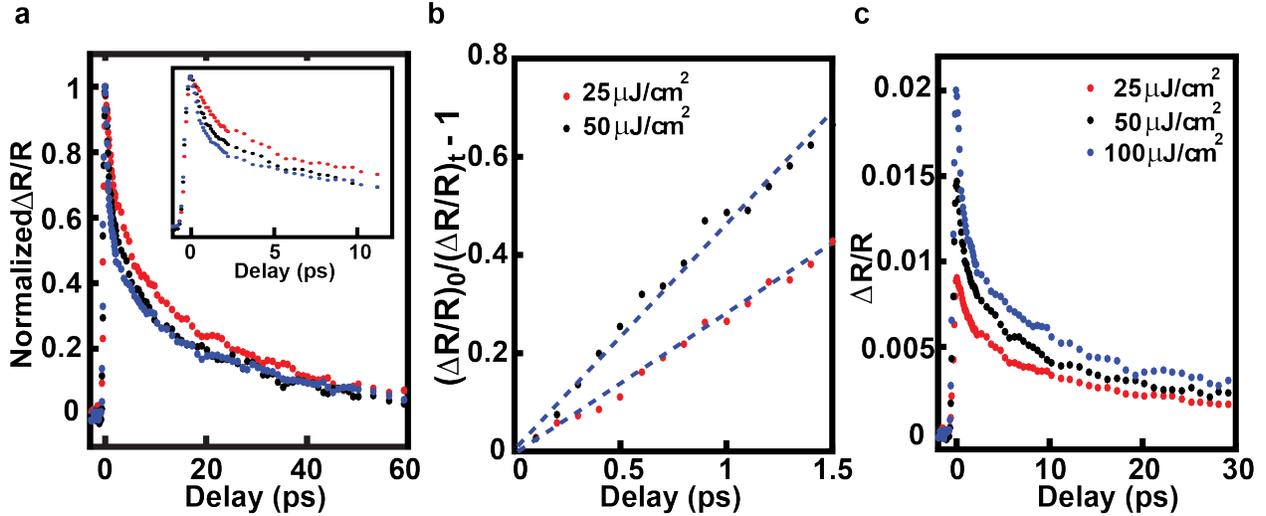

**Figure S5.** (a) Normalized differential reflection of as-grown MoS$_2$ as a function of the time delay and with different pumping fluences, 25 μJ/cm$^2$ (red), 50 μJ/cm$^2$ (black), and 100 μJ/cm$^2$ (blue). Inset: the results for the early stage of the decay. (b) The result of $(ΔR/R)_0/(ΔR/R)_t -1$ derived from the data in (a) as a function of the delay time. The result for the pumping fluence of 100 μJ/cm$^2$ is not shown for the visual convenience. (c) The non-normalized differential reflection of as-grown MoS$_2$ at pumping fluence 25 μJ/cm$^2$ (red), 50 μJ/cm$^2$ (black), and 100 μJ/cm$^2$ (blue).

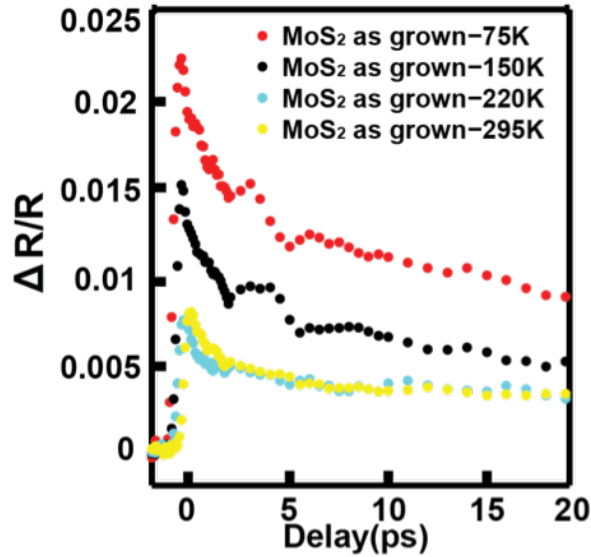

**Figure S6.** Transient reflection ΔR/R measured at the supported monolayer MoS2 on sapphire substrates at different temperatures, 75K, 150K, 220K, and 295K. The ΔR/R at the 0s delay exhibits a strong dependence on the temperature.

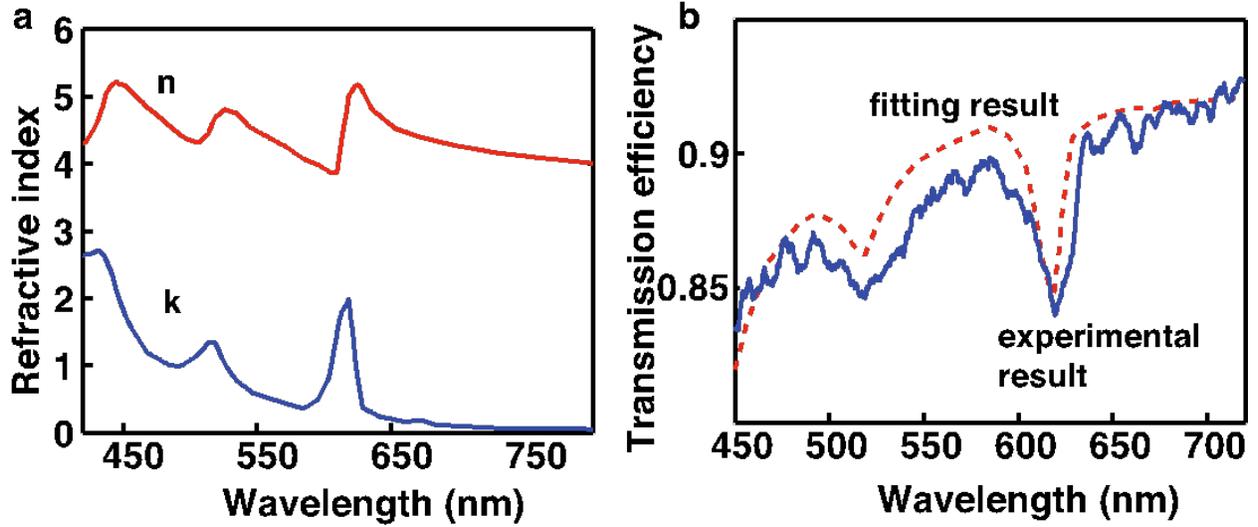

**Figure S7.** Measured refractive index of monolayer WS$_2$. (a) measured real and imaginary part of the refractive index of monolayer WS$_2$. (b) measured and fitted transmission of as-grown monolayer WS2 on sapphire substrates. The fitting results uses the refractive index given in (a).

## S1. Derivative for the population inversion in the monolayers

The rate equation for the charge density at the level 2 can be written as

$$\frac{dN_2}{dt} = -\sigma_{12}\frac{I}{h\nu_{12}}(N_2 - N_1) - \left(\frac{1}{\tau_r} + \frac{1}{\tau_{nr}}\right)N_2 - k_{ee}N_2^2 + \sigma_{13}\frac{I_p}{h\nu_{13}}N_1 \quad \text{(S1)}$$

where $\sigma_{12}$ is the stimulated emission (absorption) cross section for the light in frequency $\nu_{12}$ that matches the energy difference between the level 1 and level 2, $I$ is the photon flux at the frequency of $\nu_{12}$, $h$ is the Planck's constant, $\sigma_{13}$ is the absorption cross section for the pumping light in frequency $\nu_{13}$, and $I_p$ is the pumping intensity. By using the steady state ($dN_2/dt = 0$) and the conservation of charges ($N_2 + N_1 = N_t$, $N_t$ is the total charge density), we can find out the population inversion ($N_2 - N_1$) as

$$\Delta N = N_2 - N_1 = \frac{\sqrt{(1/\tau_r + 1/\tau_{nr} + W_p)^2 + 4k_{ee}W_p N_t} - (1/\tau_r + 1/\tau_{nr} + W_p) - k_{ee}N_t}{k_{ee}} \quad \text{(S2)}$$

and the optical gain coefficient as $\gamma = \sigma_{12}\Delta N$, where $W_p = \sigma_{13}I_p/h\nu_{13}$ representing the pumping rate. In the process of deriving eq. (S2) we assume a small-signal inversion ($I \ll I_p$) and ignore the term of $\sigma_{12}I/h\nu_{12}$.

Based on the value given in Table 1, $k_{ee}N_t/2$ is always one or two orders of magnitude larger than $1/\tau_r + 1/\tau_{nr}$ for all the supported and suspended monolayers. As a result, essentially the threshold pumping rate $W_p = k_{ee}N_t/2$ as $k_{ee}N_t/2 \gg 1/\tau_r + 1/\tau_{nr}$. And the population inversion can be simplified as

$$\Delta N = \left(\sqrt{W_p^2 + 4k_{ee}W_p N_t} - W_p\right)/k_{ee} - N_t. \quad \text{(S3)}$$